\documentclass[a4paper,12pt]{article}

\newcommand{\sect}[1]{\setcounter{equation}{0}\section{#1}}

\begin{document}
\topmargin 0pt \oddsidemargin 0mm

\renewcommand{\thefootnote}{\fnsymbol{footnote}}
\begin{titlepage}
\begin{flushright}
\end{flushright}

\vspace{5mm}
\begin{center}
{\Large \bf Hawking temperature for  constant curvature black bole
and its analogue in de Sitter Space} \vspace{12mm}

{
Rong-Gen Cai$^{1,}$\footnote{Email address: cairg@itp.ac.cn} and Yun Soo Myung$^{2,}$\footnote{Email address: ysmyung@inje.ac.kr} \\
\vspace{8mm} { \em $^1$Key Laboratory of Frontiers in Theoretical
Physics, Institute of Theoretical Physics, Chinese Academy of
Sciences,
   P.O. Box 2735, Beijing 100190, China \\
   $^2$Institute of Basic Science and School of Computer Aided Science, Inje University,
Gimhae 621-749, Korea} }
\end{center}
\vspace{5mm} \centerline{{\bf{Abstract}}}
 \vspace{5mm}
 The constant curvature (CC) black holes are higher dimensional
 generalizations of BTZ black holes. It is known that these black holes have
 the unusual topology of ${\cal M}_{D-1}\times S^1$, where $D$ is the
 spacetime dimension and ${\cal M}_{D-1}$ stands for a conformal Minkowski
 spacetime in $D-1$ dimensions. The unusual topology and time-dependence
  for the exterior of  these black holes cause some difficulties to
  derive their thermodynamic quantities.  In this work, by using globally
  embedding approach, we obtain the Hawking temperature of the
  CC black holes.  We find that the Hawking
  temperature takes  the same form when using  both the static  and
  global coordinates.  Also it is identical to the Gibbons-Hawking
  temperature of the boundary de Sitter spaces of these CC  black holes. Employing the same approach, we
  obtain the Hawking temperature for the counterparts of CC black holes in de Sitter spaces.

\end{titlepage}

\newpage
\renewcommand{\thefootnote}{\arabic{footnote}}
\setcounter{footnote}{0} \setcounter{page}{2}
\sect{Introduction}
 Over the past years, the so-called BTZ (Banados-Teitelboim-Zanelli)~\cite{BTZ} black hole solutions have
 played the important role in understanding microscopic degrees of
 freedom of black hole.  The BTZ black hole  is an exact
 solution of Einstein field equations with a negative cosmological
 constant in three dimensions. It is well known  the BTZ black hole can be
 constructed by identifying  points along the orbit of a Killing vector in
 a three dimensional anti-de Sitter (AdS) space.

 The BTZ black hole has a topology of ${\cal M}_2 \times S^1$, where $M_2$ denotes a conformal
 Minkowski space in two dimensions.  Following the same way as done in three dimensions, one can
 construct analogues of the BTZ solution,  the so-called constant curvature (CC) black holes  in higher $(D\ge 4$)
 dimensional AdS spaces~\cite{Holst,Bana1,Bana2}.  However, such
 black holes have topology of ${\cal M}_{D-1}\times S^1$ in $D$
 dimensions, which is quite different from the known topology of ${\cal M}_2
 \times S^{D-2}$ for the usual black holes in $D$ dimensions.   In addition,
  the exterior region  of these CC black holes is
 time-dependent and thus, there is no global time-like Killing
 vector~\cite{Holst}.  Because of  this,
 it is difficult to discuss Hawking radiation and
 thermodynamics associated with these   black
 holes. For example, see \cite{Mann,Ross,Hut} and references
 therein.

 On the other hand, these spacetimes are interesting examples of smooth
  time-dependent solutions. Particularly,   they are consistent background spacetimes
  for string theory at least to leading order  since they are
  vacuum solutions to Einstein field equations with a negative cosmological constant too.
  Further we note that  these
  spacetimes are time-dependent, the boundary metric is also
  time-dependent, and it is asymptotically  AdS.  Therefore, it might   open a window to
 investigate dual strong coupling field theory in the time-dependent
  backgrounds through the AdS/CFT correspondence~\cite{AdS}.
  Especially, the $D$-dimensional CC black holes have the boundary
  topology of $dS_{D-2}\times S^1$, where $dS_{D-2}$ denotes a
  $(D-2)$-dimensional de Sitter (dS) space.  Resorting the AdS/CFT
  correspondence,  these CC black holes are
  gravity duals to strong coupling conformal field theories living
  on $dS_{D-2}\times S^1$.  Finally it is observed in \cite{Cai1} that these CC
  black holes have a  close connection to the so-called ``bubbles
  of  nothing" in AdS space~\cite{Birm,BR}.  The bubbles of nothing
  were
  constructed by analytically continuing (Schwarzschild, Reissner-Nordstr\"om,
  and Kerr) black holes in AdS spaces.  The stress-energy tensor
  for dual conformal field theories to these CC
  black holes was  calculated in~\cite{Cai1,BR}.

It is well known that there is the Gibbons-Hawking temperature
$T_{\rm GH}$ for a comoving observer in a dS space~\cite{GH}. This
temperature may be viewed as the Hawking temperature $T_{\rm HK}$
associated with cosmological horizon of dS space.  A $D$-dimensional
dS space can be embedded as a hypersurface into a
$(D+1)$-dimensional Minkowski space. Then, the comoving observer in
dS space is identical to an observer with a constant acceleration in
Minkowski space. According to Davies~\cite{Davies} and
Unruh~\cite{Unruh}, an observer with a constant acceleration in
Minkowski space will see a hot bath with the  Davies-Unruh
temperature $T_{\rm DU}=a/2\pi$  where $a$ is the acceleration of
the observer.  It turns out that the Gibbons-Hawking temperature of
dS space is equivalent to the Davies-Unruh temperature of the
corresponding observer in  Minkowski space.

One decade ago, it was shown  that an observer with a constant
acceleration $a$ in dS space will detect a temperature given by
$\sqrt{a^2 +1/l^2}/2\pi$, where $l$ is the radius of the dS
space~\cite{Narn}.  This was soon generalized to the cases of dS/AdS
space by Deser and Levin~\cite{DL} with temperatures of $\sqrt{a^2
\pm 1/l^2}/2\pi$.  Further, Deser and Levin have shown that the
temperature is equivalent to the Davies-Unruh temperature for the
corresponding observer in  Minkowski space.  Further examples for
the equivalence have been shown by globally embedding curved spaces
including BTZ, Schwarzschild, Schwarzschild-AdS (dS), and
Reissner-Nordstr\"om solutions into higher dimensional Minkowski
spaces in Ref.\cite{DL2}. For more examples on the equivalence, see
\cite{Kim} and references therein.

In this work,  the ``globally embedding approach"  will be employed
to determine the Hawking temperature of CC black holes and positive
CC spaces. This approach  shows a clear way to compute the Hawking
temperature, in comparison to other methods with ambiguity to
calculate it.

 In the next
section, we show that the Hawking temperature of the constant
curvature black holes is given by $T_{\rm HK}=r_+/(2\pi l)$ using
both the static and global coordinates in AdS spaces.  Further it is
shown that the Hawking temperature is identical to the
Gibbons-Hawking temperature of the boundary dS space.  In section 3
we consider the counterparts of the CC black holes in dS spaces.
These are constant curvature (CC) spaces with the cosmological
horizon. We find that the Hawking temperature for these spaces are
given by $T_{\rm HK}=r_+/(2\pi l)$, where $r_+$ and $l$ are the
cosmological horizon radius and the radius of dS spaces,
respectively.   We give our conclusions and discussions in section
4.  In this paper, we confine ourselves to the five dimensional
space. The generalization to other dimensions is straightforward.

\sect{Hawking temperature of CC black holes}

A five dimensional AdS space is defined as the universal covering
 space of a surface obeying
\begin{equation}
\label{2eq1}
 -x_0^2 +x_1^2 +x_2^2 +x_3^2 +x_4^2-x_5^2=-l^2,
\end{equation}
where $l$ is the AdS radius. This surface has fifteen  Killing
vectors of  seven rotations and eight boosts. We consider  the boost
$\xi=(r_+/l)(x_4\partial_5+x_5\partial_4)$ with its  norm
$\xi^2=r_+^2(-x_4^2+x_5^2)/l^2$  where $r_+$ is an arbitrary real
constant.  The so-called CC black hole is constructed by identifying
points along the orbit of the Killing vector $\xi$.  Since the
starting point is the AdS, the resulting black hole has a constant
curvature as the AdS does show.  The topology of the black holes is
changed to be ${\cal M}_{4}\times S^1$, which is quite different
from the usual topology of ${\cal M}_2\times S^3$ for  five
dimensional black holes. Here ${\cal M}_n$ denotes a conformal
Minkowski space in $n$ dimensions. For more details for the
construction of the black hole, see~\cite{Bana1,Bana2}.

The CC black holes can be nicely described by using Kruskal
coordinates. For this purpose, a set of coordinates on the AdS for
the region of $\xi^2 >0$ has been introduced  in Ref.\cite{Bana1}.
The six dimensionless local coordinates $(y_i,\varphi)$ are given by
\begin{eqnarray}
\label{2eq2}
 && x_i=\frac{2ly_i}{1-y^2}, \ \ \ i=0,1,2,3 \nonumber \\
 && x_4=\frac{lr}{r_+}\sinh\left(\frac{r_+\varphi}{l}\right),
 \nonumber \\
 && x_5=\frac{lr}{r_+}\cosh\left(\frac{r_+\varphi}{l}\right),
 \end{eqnarray}
 with
 \begin{equation}
 r=r_+\frac{1+y^2}{1-y^2}, \ \ \ y^2=-y_0^2+y_1^2+y_2^2+y_3^2.
 \end{equation}
Here the allowed regions are $ -\infty <y_i < \infty$ and $-\infty
<\varphi <\infty$ with the restriction $-1 <y^2 <1$. In these
coordinates,  the boundary at $r\to \infty$ corresponds to the
hyperbolic ``ball"  which satisfies $y^2=1$.  The induced line
element can be written down
\begin{equation}
\label{2eq4}
 ds^2 = \frac{l^2(r+r_+)^2}{r_+^2}(-dy_0^2
+dy_1^2+dy_2^2+dy_3^2)
    +r^2 d\varphi^2.
\end{equation}
Obviously, the Killing vector is given by $\xi=\partial_{\varphi}$
with its norm $\xi^2=r^2$.  The black hole spacetime could be
obtained by identifying $\varphi \sim \varphi +2\pi n$, and the
topology of the black hole  takes  the form of ${\cal M}_4\times
S^1$ clearly.

On the other hand, the CC black holes can also be described by
introducing Schwarzschild coordinates. The local ``spherical"
coordinates ($t,r,\theta,\chi$) in the hyperplane $y_i$ are
\begin{eqnarray}
\label{2eq5}
 && y_0=f \sin\theta \sinh(r_+t/l),\ \ \
y_1=f\sin\theta
    \cosh(r_+t/l), \nonumber \\
 && y_2=f \cos\theta \sin\chi, \ \ \ \ \ \ y_3=f \cos\theta \cos\chi,
\end{eqnarray}
where $f=[(r-r_+)/(r+r_+)]^{1/2}$, $0\le \theta \le \pi/2$, $0\le
\chi \le 2\pi$, and $r_+ \le r< \infty$. One finds that the solution
(\ref{2eq4}) can be rewritten as
\begin{equation}
\label{2eq6}
 ds^2= l^2 N^2 d\Omega_3^2 +N^{-2}dr^2 +r^2d\varphi^2,
 \end{equation}
 where
 \begin{equation}
 \label{2eq7}
 N^2=\frac{r^2-r_+^2}{l^2}, \ \ \ d\Omega_3^2=-\sin^2\theta dt^2
 +\frac{l^2}{r_+^2}(d\theta^2 +\cos^2\theta d\chi^2).
 \end{equation}
 This is the black hole solution expressed in terms of Schwarzschild coordinates.
 Here $r=r_+$ is the location of  black hole horizon.
 In these coordinates the solution seems static. However,  we
 observe
 from (\ref{2eq6}) that the form (\ref{2eq7}) does not cover
 the whole exterior region of black hole since the difference
 of $y^2_1-y_0^2$ is required  to be positive in the region covered
 by these coordinates.  Indeed, it has been proved that there is no
 globally timelike Killing vector in this geometry~\cite{Holst}.

Now we  consider a static observer with constant $(r>r_+, \theta,
\chi, \varphi)$ in the black hole background (\ref{2eq6}). To this
observer, we find that an associated acceleration $a_5$ is given by
\begin{equation}
\label{2eq8} a_5^2= \frac{1}{l^2(r^2-r_+^2)}
\frac{1}{\sin^2\theta}\left(r^2 \sin^2\theta +r_+^2
\cos^2\theta\right).
\end{equation}
On the other hand,  the acceleration of $a_6$ for the corresponding
observer in six dimensional embedding Minkowski space is given by
\begin{equation}
\label{2eq9} a_6^{-2} = x_1^2-x_0^2=
\frac{l^2(r^2-r_+^2)}{r_+^2}\sin^2\theta.
\end{equation}
It is easy to check that these two accelerations obey the relation
\begin{equation}
\label{2eq10} a_6^2 = -\frac{1}{l^2}+a^2_5.
\end{equation}
This shows that the Davies-Unruh temperature for the local observer
in six dimensional Minkowski space is
\begin{equation}
T_{\rm DU}=\frac{a_6}{2\pi}= \frac{r_+}{2\pi
l\sqrt{r^2-r_+^2}}\frac{1}{\sin\theta}.
\end{equation}
We note that the redshift factor of $\sqrt{-g_{00}}=l N\sin\theta$
for the black hole (\ref{2eq6}) is necessary to define the Hawking
temperature. Hence we conclude that the Hawking temperature of the
CC black hole is
\begin{equation}
\label{2eq12}
 T_{\rm HK}=\sqrt{-g_{00}}~ T_{\rm DU}= \frac{r_+}{2\pi l}.
\end{equation}
 We notice that the Hawking
temperature $T_{\rm HK}$  is consistent with the inverse period of
the Euclidean time derived from the solution (\ref{2eq7}).

As the case in four dimensions~\cite{Bana2}, there is another set of
coordinates covering  the whole  exterior of the Minkowskian black
hole geometry as~\cite{Cai1}
\begin{eqnarray}
\label{2eq13}
 && y_0 = f\sinh(r_+t/l), \ \ \ y_1= f\cos\theta \cosh(r_+t/l),
 \nonumber \\
  && y_2= f\sin\theta \cos\chi \cosh(r_+t/l), \ \ \
  y_3=f\sin\theta\sin\chi \cosh(r_+t/l),
  \end{eqnarray}
  where $f$ is given by (\ref{2eq5}) and the allowed regions are  $ 0\le \theta \le \pi$, $r_+ \le
  r<\infty$, and $0 \le \chi \le 2\pi $. In  these
  coordinates, the solution can be expressed as
  \begin{equation}
  \label{2eq14}
  ds^2=N^2l^2 d\Omega_3^2+ N^{-2}dr^2 +r^2d\varphi^2,
  \end{equation}
  where $N^2=(r^2-r_+^2)/l^2$ and
  \begin{equation}
  \label{2eq15}
  d\Omega_3^2=-dt^2 +\frac{l^2}{r_+^2} \cosh^2(r_+t/l)(d\theta^2
  +\sin^2\theta d\chi^2).
  \end{equation}
 The time-dependence of the solution
  is manifest in this coordinate system.  We introduce  a static observer located at  constant
  $r>r_+$,  $\varphi$ and $\chi$, but $\theta=0$ due to the
  spherical symmetry of the solution~\cite{DL}. Here, we find that
  the acceleration $a_5$ associated with the observer is
  \begin{equation}
\label{2eq16} a_5^2= \frac{r^2}{l^2 (r^2-r_+^2)},
\end{equation}
while the acceleration $a_6$ of  the corresponding observer in six
dimensional Minkowski space is given by
 \begin{equation}
  a^{-2}_6= x_1^2-x_0^2=\frac{r_+^2}{l^2(r^2-r_+^2)}.
  \end{equation}
  They satisfy  the relation (\ref{2eq10}) too. In this case, the Davies-Unruh
  temperature is given by
  \begin{equation}
  T_{\rm DU}= \frac{a_6}{2\pi}= \frac{r_+}{2\pi l \sqrt{r^2-r_+^2}}.
  \end{equation}
  Considering the redshift factor of $\sqrt{-g_{00}}$,  we get the Hawking temperature of the black hole in the
 line element of  (\ref{2eq14}) as
  \begin{equation}
  \label{2eq19}
  T_{\rm HK}=\frac{r_+}{2\pi l}.
  \end{equation}
Thus we have obtained the Hawking temperature of the CC black hole
by employing  globally embedding approach combined  with the
Davies-Unruh temperature in six dimensional Minkowski space. The
Hawking temperatures (\ref{2eq12}) and (\ref{2eq19}) are our main
results.

Here some remarks are in order. First, in  general, Hawking
temperature of black hole depends on coordinates used to calculate
it. That is, the Hawking temperature may be different when using
different coordinates, even for the  same black hole. In our case,
we obtained the same Hawking temperature for the CC black hole even
when used the different coordinate systems  (\ref{2eq6}) and
(\ref{2eq14}). Second, we mention that  the Hawking temperature
(\ref{2eq12}) is the same as the inverse period of the Euclidean
time for the Euclidean sector of the solution (\ref{2eq6}). However,
when used the coordinates (\ref{2eq15}), the Hawking temperature is
no longer the same as the inverse period of the Euclidean time.   In
order to see this, let us consider carefully  the Euclidean sector
of the black hole solution which can be obtained by replacing the
time
 $t$ by $ -i( \tau + \pi l /(2 r_+))$ in (\ref{2eq15}). In this case,
 $d\Omega_3^2$ becomes
 \begin{equation}
 \label{2eq16}
 d\Omega_3^2=d\tau^2
 +\frac{l^2}{r_+^2}\sin^2(r_+\tau/l)(d\theta^2+\sin^2\theta d\chi^2).
 \end{equation}
In order that  $d\Omega_3^2$ be  a regular three-sphere, $\tau$ must
have the period of $\tau \sim \tau +\tilde \beta $ with
\begin{equation}
 \tilde \beta =\frac{\pi l}{r_+}.
 \end{equation}
Clearly this is not the inverse of Hawking temperature.  This shows
that the Euclidean method does not always provide a correct Hawking
temperature of CC black holes.  However, using  both coordinates
(\ref{2eq5}) and (\ref{2eq13}), the Euclidean time $\tau =i t$
obtained by Wick rotation leads to the fact  that it has a
periodicity with period $2\pi l/r_+$, which gives a correct Hawking
temperature of the black hole. This may be related to  the issue of
the factor 2 in~\cite{Yale}.

Finally, we observe  from (\ref{2eq14}) that the black hole solution
has a boundary topology $dS_3 \times S^1$  at $r= \infty$. The three
dimensional de Sitter space $dS_3$ has a Hubble constant $H=r_+/l$.
It is well known that for a de Sitter space with a Hubble constant
$H$, there is the Gibbons-Hawking temperature $T=H/2\pi$ for a
comving observer. We find that the Gibbons-Hawking temperature in
our case is identical to the Hawking temperature of the  CC black
hole
\begin{equation}
T_{\rm GH}= \frac{H}{2\pi}=\frac{r_+}{2\pi l}=T_{\rm HK}.
\end{equation}

\sect{Hawking temperature of a positive CC space}

In this section we consider the analogue of the CC black hole in dS
space. This space is constructed by identifying points along the
orbit of a Killing vector in dS space. In fact, this space is a
generalization of  the three-dimensional Schwarzschild-de Sitter
solution in higher dimensions. This space has a cosmological event
horizon, and  its topology  is   ${\cal M}_{D-1}\times S^1$ where
${\cal M}_{D-1}$ denotes a $(D-1)$-dimensional conformal Minkowski
spacetime. Such space was constructed in Ref.\cite{Cai}.

As the case with a negative cosmological constant, we consider a
five dimensional de Sitter space, which can be viewed as a
hypersurface embedded into a six dimensional Minkowski space,
satisfying
\begin{equation}
\label{3eq1}
 -x_0^2+x_1^2 +x_2^2 +x_3^2 +x_4^2 +x_5^2 =l^2
 \end{equation}
 with $l$ the radius of the dS space.  This dS space has
 fifteen Killing vectors of five boosts and ten rotations.
 We consider a rotational  Killing vector $\xi =(r_+/l)(x_4\partial _5
 -x_5\partial_4)$ with its norm $\xi^2 = r_+^2/l^2 (x_4^2 +x_5^2)$
 where $r_+$ is an arbitrary real constant. Identifying points along the orbit of a Killing
 vector $\xi$, another one-dimensional manifold becomes compact and it is isomorphic
 to $S^1$. Thus, we  obtain a spacetime of topology ${\cal M}^4 \times S^1$ with cosmological horizon.
  For details of the construction of the space, see \cite{Cai}.

We can describe the spacetime in the region with $0\le \xi^2 \le
r_+^2$ by introducing six dimensionless local coordinates $(y_i,
\phi)$,
\begin{eqnarray}
\label{3eq2}
 && x_i=\frac{2ly_i}{1+y^2}, \ \ \ i=0,1,2,3 \nonumber \\
 && x_4=\frac{lr}{r_+}\sin\left(\frac{r_+\phi}{l}\right),
 \nonumber \\
 && x_5=\frac{lr}{r_+}\cos\left(\frac{r_+\phi}{l}\right),
 \end{eqnarray}
 where
 \begin{equation}
 r=r_+\frac{1-y^2}{1+y^2}, \ \ \ y^2=-y_0^2+y_1^2 +y_2^2+y_3^2.
 \end{equation}
Here the allowed regions are  $-\infty <y_i <+\infty$  and $-\infty
<\phi <+\infty$ with the restriction $-1<y^2<1$ to have a positive
$r$. In the coordinates (\ref{3eq2}), the induced line element  is
\begin{equation}
\label{3eq4}
 ds^2=\frac{l^2(r+r_+)^2}{r_+^2}(-dy_0^2+dy_1^2+dy_2^2+dy_3^2)
 +r^2 d\phi^2,
 \end{equation}
 which is the same form as the case of a  negative constant
 curvature~\cite{Bana1}. However, it is noted
 that the coordinates (\ref{3eq2}) and the definition of $r$
 differ from those in the CC
 black holes.
In this coordinate system, it is evident that the Killing vector is
  $\xi =\partial_{\phi}$ with norm $\xi^2=r^2$.  Imposing the
  identification $\phi \sim \phi +2\pi n$, the solution has
  the topology ${\cal M}_4 \times S^1$.

 We now introduce the Schwarzschild coordinates to describe the
solution. Using local ``spherical" coordinates $(t,r,\theta,\chi)$
defined as
\begin{eqnarray}
\label{3eq5}
 && y_0=f\sin\theta\sinh(r_+t/l), \ ~~~~~ y_1=f\sin\theta
 \cosh(r_+t/l), \nonumber \\
 && y_2=f\cos\theta \sin\chi, \ ~~~~~ y_3=f\cos\theta \cos\chi,
 \end{eqnarray}
 where $f=[(r_+-r)/(r+r_+)]^{1/2}$, and the allowed coordinate ranges are
 $ 0< \theta <\pi/2$, $0<\chi <2\pi$, and $0<r<r_+$.
 The line element  can be expressed as
 \begin{equation}
 \label{3eq6}
 ds^2 =l^2 N^2d\Omega_3^2 +N^{-2}dr^2 +r^2d\phi^2.
 \end{equation}
 Here $N^2=(r_+^2-r^2)/l^2$ and
 \begin{equation}
 \label{3eq7}
 d\Omega_3^2= -\sin^2\theta dt^2 +\frac{l^2}{r_+^2}(d\theta^2
 +\cos^2\theta d\chi^2).
 \end{equation}
 Clearly the location of $r=r_+$ represents a cosmological horizon.
 This solution  is  the counterpart of a five dimensional CC black hole described in the previous section.
 The only difference is that $N^2=(r^2-r_+^2)/l^2$ is
 replaced by $N^2=(r_+^2-r^2)/l^2$ here.  Further, in three dimensions, the
 corresponding induced line element takes the form
 \begin{equation}
 \label{3eq8}
 ds^2=-(r_+^2-r^2)dt^2 +\frac{l^2}{r_+^2-r^2}dr^2 +r^2d\phi^2,
 \end{equation}
 After a suitable rescaling of coordinates, it can be transformed
 to  three dimensional Schwarzschild-de Sitter
 solution~\cite{Park}.  In this sense, the solution (\ref{3eq6}) can
 be viewed as an analogue of the three dimensional Schwarzschild-de Sitter
 solution in five dimensions.

  The solution (\ref{3eq6}) seems to be static, but it does not
 cover the whole region within the cosmological horizon. It can be seen from
 the definition of coordinates (\ref{3eq5}) because they must obey the
 constraint: $y_1^2-y_0^2=f^2\cos^2\theta \ge 0$.
  Considering a static observer located at constant $(r<r_+, \theta,
 \chi$) in the background (\ref{3eq6}), we find that the
 static observer has a constant acceleration $a_5$ as
 \begin{equation}
 \label{3eq9}
 a_5^2 =\frac{1}{l^2 (r_+^2-r^2) \sin^2\theta}\left(
 r^2\sin^2\theta+ r_+^2\cos^2\theta\right),
 \end{equation}
 while the observer in six dimensional Minkowski space
  has a constant acceleration
 $a_6$ as
 \begin{equation}
 a_6^{-2} = x_1^2-x_0^2= \frac{l^2 (r_+^2-r^2)}{r_+^2}\sin^2\theta.
 \end{equation}
 These two accelerations are related to each other as
 \begin{equation}
 a_6^2 =1/l^2 +a_5^2.
 \end{equation}
 According to Davies and Unruh, the observer has a temperature as
 \begin{equation}
 T_{\rm DU}= \frac{a_6}{2\pi}=\frac{r_+}{2\pi l\sqrt{r_+^2-r^2}\sin\theta}.
 \end{equation}
Taking into account  the redshift factor $\sqrt{-g_{00}}$ of the
observer, one has the Hawking temperature  as
\begin{equation} \label{3eq13}
T_{\rm HK}=\frac{r_+}{2\pi l}.
\end{equation}

On the other hand, one has  another set of coordinates which covers
the whole region within the cosmological horizon,
\begin{eqnarray}
\label{3eq14}
 && y_0=f\sinh(r_+t/l), \ ~~~~~~ y_1=f\cos\theta
\cosh(r_+t/l),
\nonumber \\
&& y_2=f\sin\theta \cos\chi \cosh(r_+t/l), \ ~~~~~~ y_3=f\sin\theta
\sin\chi \cosh(r_+t/l).
\end{eqnarray}
In this case, the line element  is described by
\begin{equation}
\label{3eq15}
 ds^2=l^2N^2\tilde {d\Omega_3^2} +N^{-2}dr^2 +r^2d\phi^2,
 \end{equation}
 where $N^2=(r_+^2-r^2)/l^2$ and
 \begin{equation}
 \label{3eq16}
 \tilde {d\Omega_3^2}=-dt^2
 +\frac{l^2}{r_+^2}\cosh^2(r_+t/l)(d\theta^2+\sin^2\theta
 d\chi^2).
\end{equation}
We consider a static observer located at constant position of
$(r<r_+, \chi, \varphi)$ and $\theta=0$. For such an observer, we
have a constant acceleration $a_5$ as
\begin{equation}
a_5^2= \frac{r^2}{l^2(r^2_+-r^2)}.
\end{equation}
In six dimensional Minkowski space, the acceleration $a_6$
associated with the corresponding observer takes the form
\begin{equation}
a_6^{-2}=x_1^2-x_0^2= \frac{l^2(r_+^2-r^2)}{r_+^2}.
\end{equation}
We check that they obey the relation $a_6^2 =a_5^2+1/l^2$. We
conclude that the observer has the Davies-Unruh temperature
\begin{equation}
T_{\rm DU}=\frac{a_6}{2\pi}= \frac{r_+}{2\pi l\sqrt{r_+^2-r^2}}.
\end{equation}
Considering  the redshift factor for the observer in the background
(\ref{3eq15}), we have the Hawking temperature observed as
\begin{equation}
  T_{\rm HK}=\frac{r_+}{2\pi l}.
  \end{equation}
Consequently,  we find  the same Hawking temperature as
(\ref{3eq13}) obtained  when using the coordinates (\ref{3eq6}).

 \sect{Conclusions and Discussions}

  A $D$-dimensional CC black hole has unusual topological
  structure ${\cal M}_{D-1}\times S^1$ and there is no globally
  timelike Killing vector in the geometry of the black hole.
  Hence it was  quite difficult to discuss thermodynamic properties
   and Hawking temperature associated with this black hole.

   For
   example,
  Banados has  considered a five dimensional rotating CC black hole and embedded it into a Chern-Simons
  supergravity theory~\cite{Bana1}. By computing related conserved charges, it
  was shown that the black hole mass is proportional to the
  product of outer horizon $r_+$ and inner horizon $r_-$, while
  the angular  momentum is proportional to the sum of two horizons.
 In this case, the entropy of black hole is found to be proportional not to
 the outer horizon $r_+$ but
  the inner horizon $r_-$.   This
  approach has two drawbacks. One is that the result cannot be
  degenerated to the non-rotating case. The other   is that it cannot be
  generalized to other dimensions.

   Creighton and Mann have considered the quasilocal
   thermodynamics of a four dimensional CC black
   hole in general relativity by computing   thermodynamic
   quantities at a finite boundary which encloses the black hole~\cite{Mann}.
   They have shown that the entropy is not associated
   with the event horizon, but the Killing horizon of a static
   observer which is tangent to the event horizon of the black
   hole.  The quasilocal energy density [see (11) of \cite{Mann}]
   is negative.

   In this work, we have derived Hawking temperature of CC  black holes by employing the globally embedding approach since these
   black holes can be embedded into higher dimensional Minkowski space.
   We found that the Hawking temperature of  CC black holes is given by $r_+/2\pi l$ when using  both static
   and global coordinates.  Here $r_+$ and $l$ are
   black hole horizon and the radius of AdS space. Furthermore we
   found that the Hawking temperature is also identical to the
   Gibbons-Hawking temperature of the boundary dS space of the
   CC black holes.  Importantly, we mention that the Hawking
   temperature obtained in this work is the same as that obtained  from
   semi-classical tunneling method~\cite{Yale}.  It turns out that
   the globally embedding technique is powerful to
   determine the Hawking temperature of CC black
   hole without any ambiguity.  Using the same approach,   we also obtained Hawking temperature
   of a  positive CC space  which is
   counterpart of CC black hole in dS space.

   Finally, we comment that those solutions including CC black holes and positive CC spaces
   depend on an arbitrary real constant $r_+$.
   The $r_+$-dependence can be made disappear by rescaling
   coordinates.   In this case, the Hawking temperature
  is given by  $1/2\pi l$.

\section*{Acknowledgments}
This work was initiated  during the APCTP joint focus program:
frontiers of black hole physics, Dec. 6-17, Pohang, Korea and Inje
workshop on gravity and numerical relativity during Dec. 16-18,
Busan, Korea, the warm hospitality extended to the authors in both
places are grateful. RGC thanks S.P. Kim for useful discussions
during APCTP focus program. This work was supported in part by a
grant from Chinese Academy of Sciences and  in part by the
National Natural Science Foundation of China under Grant Nos.
10821504, 10975168 and 11035008, and by the Ministry of Science
and Technology of China under Grant No. 2010CB833004.  YSM was
supported by the National Research Foundation of Korea (NRF) grant
funded by the Korea government (MEST) (No.2009-0062869).


\end{document}